\documentclass[epj]{svjour}
\usepackage{graphics}
\begin{document}
\renewcommand{\textfraction}{0.00000000001}
\renewcommand{\floatpagefraction}{1.0}

\title{Photoproduction of $\pi^0$-mesons from nuclei}
\author{B. Krusche\inst{1},
J.~Lehr\inst{2},
J.~Ahrens\inst{3},
J.R.M.~Annand\inst{4},
R.~Beck\inst{3},
F.~Bloch\inst{1},
L.S.~Fog\inst{4},
D.~Hornidge\inst{3},
S.~Janssen\inst{5},
M.~Kotulla\inst{1},
J.C.~McGeorge\inst{4},
I.J.D.~MacGregor\inst{4},
J.~Messchendorp\inst{5},
V.~Metag\inst{5},
U.~Mosel\inst{2},
R.~Novotny\inst{5},
R.O.~Owens\inst{4},
M.~Pfeiffer\inst{5},
R.~Sanderson\inst{4},
S.~Schadmand\inst{5},
D.P.~Watts\inst{4}
}                     
\offprints{Bernd.Krusche@unibas.ch}          %
\institute{Department of Physics and Astronomy, University of Basel,
           Ch-4056 Basel, Switzerland \and
	   Institut f\"ur Theoretische Physik I, Universit\"at Giessen, D-35392
           Giessen, Germany \and
	   Institut f\"ur Kernphysik, Johannes-Gutenberg-Universit\"at Mainz,
           D-55099 Mainz, Germany \and
	   Department of Physics and Astronomy, University of Glasgow, Glasgow
	   G12 8QQ, UK \and 
	   II. Physikalisches Institut, Universit\"at Giessen, D-35392
           Giessen, Germany 
           }
\date{Received: date / Revised version: date}
%
\authorrunning{B. Krusche et al.}
\titlerunning{Neutral pion photoproduction from nuclei}

\abstract{
Photoproduction of neutral pions from nuclei (carbon, calcium, niobium, lead) 
has been studied for incident photon energies from 200 MeV to 800 MeV with 
the TAPS detector using the Glasgow photon tagging spectrometer at the Mainz 
MAMI accelerator. Data were obtained for the inclusive photoproduction of 
neutral pions and the partial channels of quasifree single $\pi^0$, double 
$\pi^0$, and $\pi^0\pi^{\pm}$ photoproduction. They have been analyzed in terms
of the in-medium behavior of nucleon resonances and the pion - nucleus 
interaction. They are compared to earlier measurements from the deuteron and 
to the predictions of a Boltzmann-Uehling-Uhlenbeck (BUU) transport model
for photon induced pion production from nuclei.
} 
\maketitle

\section{Introduction}
\label{sec:1}
The study of meson photoproduction from nuclei is motivated by two
strongly interrelated aspects, namely possible in-medium modifications of
hadrons and the meson - nucleus interaction. The in-medium properties of 
mesons and nucleon resonances is a topic which is hotly debated. 
However, with the exception of the $\Delta$-isobar, experimental results are 
still scarce and partly contradictory. In-medium modifications of nucleon 
resonances may arise from many different effects. The most trivial one is 
the broadening of the excitation functions due to Fermi motion of the bound 
nucleons. Pauli-blocking of final states decreases the decay widths but this
is counterbalanced by collisional broadening due to resonance-nucleon
scattering. In particular for the D$_{13}$(1520) it has been found in
\cite{Peters_98,Korpa_03,Post_03} that a large collisional broadening can
be generated from $\pi$ and $\rho$ exchange and that the in-medium width
is sensitive to modifications of the spectral functions of these mesons.
Besides such conventional effects arising from meson-nucleon scattering,
additional mechanisms \cite{Brown_91,Rapp_00} could lead to a shift of
spectral strength of the $\rho$ meson to smaller invariant masses, thus
further enhancing the width of the D$_{13}$.

\begin{figure}[th]
\resizebox{0.5\textwidth}{!}{%
  \includegraphics{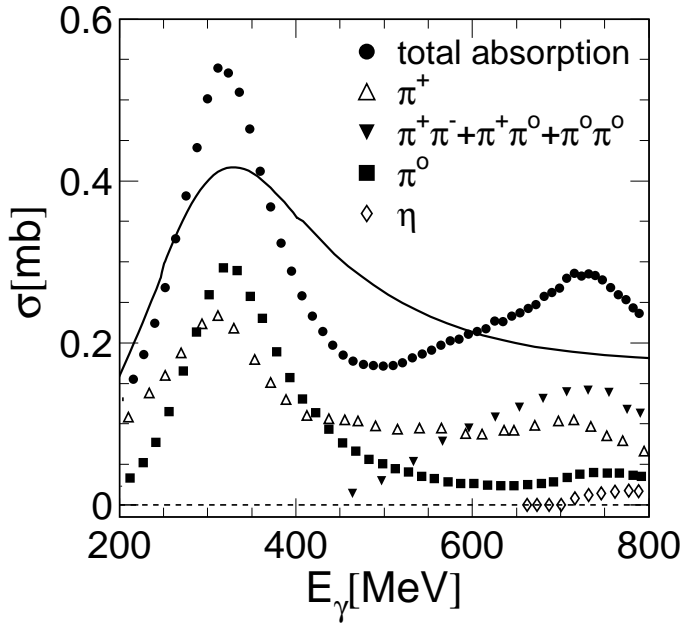}   
}
\caption{Total photoabsorption on the proton and on nuclei. 
The symbols show the total cross section and the  
decomposition into partial reaction channels for the proton. 
\cite{Buechler_94,Braghieri_95,Krusche_95,MacCormick_96,Haerter_97,Krusche_99}.
Solid curve: average of the total photoabsorption 
cross sections for nuclei scaled to their mass numbers \cite{Bianchi_94}.
}
\label{fig:proton_split}       
\end{figure}

Photon induced reactions are particularly well suited to study such effects
since photons probe the entire nuclear volume. Reactions induced by strongly 
interacting particles like pions are always restricted to the surface region
and thus probe a lower effective density. The first experimental investigation 
of the nuclear response in the resonance region to photons was done with 
total photoabsorption measurements \cite{Frommhold_94,Bianchi_94} for nuclei 
ranging from $^7$Li to $^{238}$U. Total photoabsorption has the advantage that 
no final state interaction effects (FSI) must be accounted for so that indeed 
the entire nuclear volume is probed. As a consequence, when scaled by the 
atomic mass number, the total cross sections from all nuclei fall practically 
on the same curve (`universal curve'). This curve shows a clear signal for 
the $\Delta$ resonance but no evidence of the bumps observed for the free 
nucleon in the second and third resonance regions 
(see fig. \ref{fig:proton_split}). This result of the 
experiments was surprising and is even today not fully understood.  

The situation is complicated by the already complex structure of the resonance 
peak for the free proton. The resonance consists of a superposition of 
reaction channels which differ in their energy dependence
(see fig. \ref{fig:proton_split}). Most of the rise 
of the cross section to the maximum around 700 MeV comes from double pion 
production. The interpretation of double pion photoproduction is 
complicated (see \cite{Krusche_03} for a recent overview), but it is known 
that contributions, which are not related to the excitation of resonances in 
the second resonance region, are significant. In particular, 
$\Delta$-Kroll-Rudermann and $\Delta$ pion-pole terms dominate the $\pi^+\pi^-$
final state which has the largest cross section of all double pion channels. 
Gomez Tejedor and Oset \cite{Gomez_96} have pointed out that for this channel 
the peaking of the cross section around 700 MeV is related to an interference 
between the leading $\Delta$-Kroll-Rudermann term and the sequential decay of 
the D$_{13}$ resonance via D$_{13}\rightarrow\Delta\pi$. Hirata et al. 
\cite{Hirata_98} have subsequently argued that the change of this interference 
effect in the nuclear medium is one of the most important reasons for the 
suppression of the bump structure.

Inclusive reactions like total photoabsorption alone do not allow
a detailed investigation of such effects. A study of the partial reaction
channels is desirable. The experimental identification of exclusive final 
states is more involved and FSI effects must be accounted 
for. The interpretation of exclusive measurements therefore always needs 
models which account for the trivial in-medium and FSI effects like absorption 
of mesons and propagation of mesons and resonances through nuclear matter. 
On the other hand, as a by-product, the analysis of the FSI effects enables 
the study of meson-nucleus interactions.    

During the last few years, in a series of experiments with the TAPS detector 
at MAMI in Mainz, we have studied exclusive photoproduction reactions from 
heavy nuclei which can be related to the excitation of certain resonances or 
production mechanisms. The most specific reaction in this respect is the
photoproduction of $\eta$-mesons which proceeds in the second resonance 
region almost entirely through the excitation of the S$_{11}$(1535) 
resonance \cite{Krusche_95,Krusche_97}. In this case no unexplained
depletion of the in-medium strength was found \cite{Roebig_96},
which is in line with theoretical findings \cite{Post_03,Inoue_02}
that the change of the S$_{11}$ self energy in the medium is rather small. The 
data were in excellent agreement with model calculations that take the trivial 
in-medium effects and FSI into account \cite{Effenberger_97,Carrasco_93}.
A similar result was later found by Yorita et al. \cite{Yorita_00} who 
measured the $\eta$ excitation function to somewhat higher incident photon 
energies. In a recent study \cite{Lehr_03}, it was shown that the data 
could be described over the full energy range by applying a momentum dependent 
S$_{11}$ nuclear potential. An attempt to study the in-medium properties of 
the D$_{13}$ resonance was undertaken with a measurement of quasifree single 
$\pi^0$ photoproduction \cite{Krusche_01}. Again in contrast to total 
photoabsorption the second resonance bump was clearly visible. 
On the free nucleon, the structure in this reaction is almost entirely
due to the D$_{13}$ resonance. A comparison to deuteron data showed no
broadening in nuclei and provided no evidence for an increased in-medium
width of the D$_{13}$. On the other hand, theoretical models 
\cite{Post_03,Korpa_03} predict such a modification. However, as discussed 
below, exclusive reaction channels are dominated by the nuclear surface region
where in-medium effects are smaller. This is mostly due to FSI effects.
Furthermore, as discussed in \cite{Lehr_01}, even in the absence of FSI, 
resonance broadening effects are suppressed in reactions which 
do not contribute to the broadening. The reason is, that when the total width
$\Gamma$ of a resonance increases as function of the nuclear density $\rho$
(e.g. due to collisional broadening), the branching ratio 
$b_{f}=\Gamma_{f}/\Gamma$ for any other decay channel $f$, which does not
contribute to the broadening, decreases as function of density. Due to this
effect, contributions from the broadened resonance to such exclusive final 
states are suppressed in the average over the nuclear volume. 
  
In the present paper we collect the results for all reaction channels
which involve $\pi^0$ photoproduction from nuclei. Data have been obtained 
for the inclusive reaction, where inclusive means that all final states with 
at least one $\pi^0$ meson were accepted. Furthermore, the quasifree 
contributions of single $\pi^0$, double $\pi^0$, and $\pi^0\pi^{\pm}$ 
photoproduction have been determined. The data are compared to the results 
of a semi-classical Boltzmann-Uehling-Uhlenbeck (BUU) transport model 
\cite{Effenberger_99,Lehr_00} which accounts for medium effects such as Fermi 
motion, nuclear binding, Pauli blocking, resonance broadening and includes a 
realistic description of the FSI. The treatment follows the trajectories of 
the particles through the nucleus and therefore yields additional 
information not accessible to experiment.
  
\section{Experimental setup}
\label{sec:2}

The experiments were carried out at the Glasgow tagged photon facility
\cite{Anthony_91} installed at the Mainz MAMI accelerator \cite{Walcher_90}.
The tagged photon facility uses bremsstrahlung photons produced with the 
855 MeV electron beam in a radiator foil. The scattered electrons are 
detected in the focal plane of the tagging spectrometer with an energy
resolution of $\approx$2 MeV. The experiments covered the photon energy 
range from 200 - 800 MeV. Solid targets with thicknesses of 25 mm (C), 
10 mm (Ca), 2.1 mm (Nb), and 0.47 mm (Pb), corresponding to roughly 0.1 
radiation lengths were irradiated. The diameter of the targets was 50 mm and 
the beam spot size at target position was approximately 30 mm. The neutral 
pions were detected, via their two photon decays, with the electromagnetic 
calorimeter TAPS \cite{Novotny_91,Gabler_94}.
The individual modules of the calorimeter are hexagonally shaped BaF$_2$ 
crystals of 25 cm length with an inner diameter of 5.9 cm. The inclusive and 
single $\pi^0$ photoproduction data were obtained with the setup discussed 
in \cite{Krusche_99}.
\begin{figure}[t]
\resizebox{0.49\textwidth}{!}{%
  \includegraphics{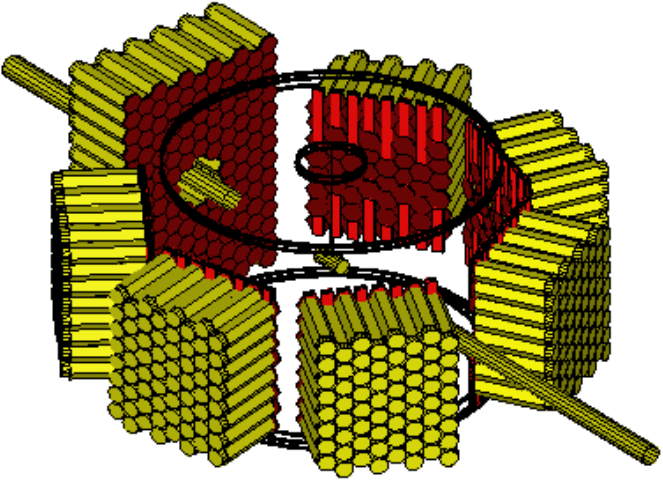} 
 
}
\caption{Setup of the TAPS detector at the Mainz MAMI accelerator.
The beam entered the target chamber from the lower right edge. 
}
\label{fig:taps}       
\end{figure}
The BaF$_2$ modules were arranged in 5 blocks, each 
containing 8$\times$8 modules. With the exception of the backward block, 
all modules were equipped with individual plastic veto detectors.
The blocks were placed in one plane around the target at a distance 
of 55 cm at polar angles of $\pm 38^o$, $\pm 88^o$, and $+133^o$. 
The double pion production reactions were studied with an upgraded version
of the TAPS detector covering a larger solid angle (see fig. \ref{fig:taps}).
In this setup 6 blocks with 64 modules and a forward wall with 138 BaF$_2$
modules were used. The blocks were placed 55 cm away from the targets at
polar angles of $\pm 54^o$, $\pm 103^o$, and $\pm 153^o$,  while the forward 
wall was placed 60 cm away from the target center. 

\section{Data analysis}
\label{sec:3}

\begin{figure}[h]
\centerline{
\resizebox{0.49\textwidth}{!}{%
  \includegraphics{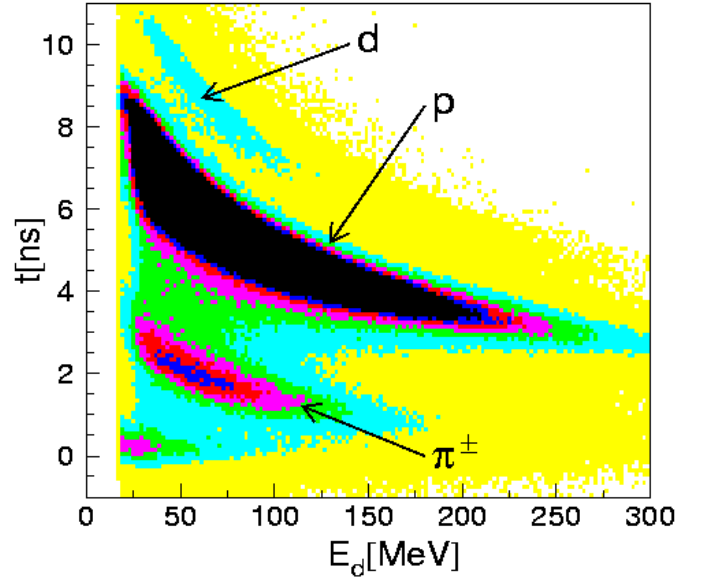} 
}}
\caption{Time-of-flight versus the energy deposited in the BaF$_{2}$ detectors. 
Three bands corresponding to deuterons, protons, and charged pions are visible
(photons already eliminated).}
\label{fig:tofvse}       
\end{figure}

The discrimination between photons and particles was done with the aid of the
plastic veto detectors, a time-of-flight analysis, and a pulse shape analysis
of the signals from the BaF$_2$ scintillators in the same way as discussed in
\cite{Krusche_99}. Neutral pions were identified with a standard 
invariant mass analysis of photon pairs. In addition to the analysis in 
\cite{Krusche_99}, charged pions for the extraction of the $\pi^0\pi^{\pm}$ 
cross section were identified from time-of-flight versus energy plots.
A typical example for this analysis is shown in figure \ref{fig:tofvse}.
Photons have already been eliminated by the methods mentioned above
and only charged hits, with an identified $\pi^0$ meson in coincidence,
are included.
Background from random coincidences between TAPS and the tagging spectrometer
was determined by shifting the time-of-flight cut away from the prompt
peak. The absolute normalization of the cross sections was done as
described in detail in \cite{Krusche_99} for the measurement with 
the deuterium target.

\subsection{Inclusive $\pi^0$ photoproduction}
\label{ssec:ana_incl}
All events with at least one $\pi^0$ identified via the invariant mass
analysis were accepted without any further kinematic cuts. Combinatorial 
background from pairs of photons originating from the decay of different 
pions was fitted and subtracted in the invariant mass spectra (see ref.
\cite{Krusche_99} for typical invariant mass spectra). For differential 
cross sections this was done separately for each bin of the respective 
observable.

\begin{figure}[h]
\resizebox{0.50\textwidth}{!}{%
  \includegraphics{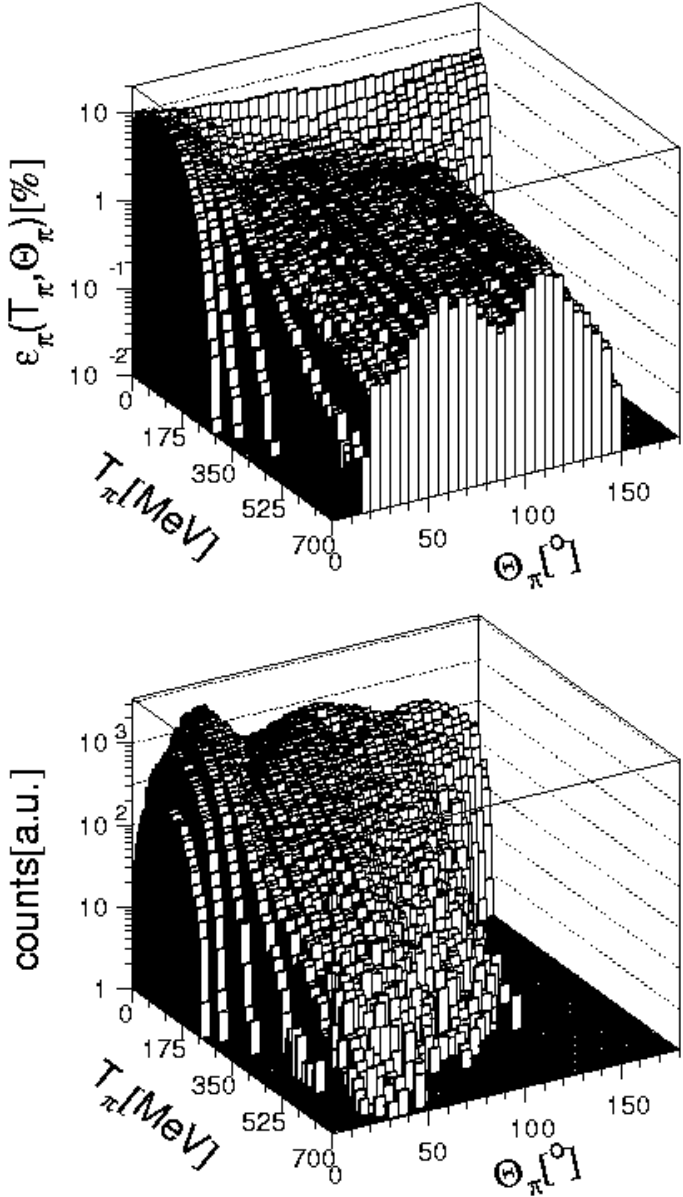}   
}
\caption{Detector acceptance for $\pi^0$-mesons as a function of the 
laboratory polar angle and the laboratory kinetic energy of the mesons. 
Upper part: efficiency calculated with a Monte Carlo simulation. 
Lower part: measured distribution of pions, not corrected for acceptance, 
for incident photon energies from 200 - 800 MeV.
}
\label{fig:effi}       
\end{figure}

The energy and angle dependent detection efficiency of the TAPS detector
was determined with Monte Carlo simulations, carried out with the GEANT3
code \cite{Brun_86}. It was previously shown that such simulations reproduce
the photon response of TAPS precisely \cite{Gabler_94}. Inclusive $\pi^0$
production is a complicated mixture of different reaction channels. With the 
exception of coherent single $\pi^0$ production all channels have at least a 
three-body final state and all are subject to final state interaction
processes. Consequently, no reliable models exist which could be used as
event generators for the Monte Carlo simulations. However, the detection
efficiency $\epsilon (\Theta_{\pi^0},T_{\pi^0})$ can only depend on the 
laboratory polar angle $\Theta_{\pi^0}$ and the laboratory kinetic energy 
$T_{\pi^0}$ of the pions (the dependence on the azimuthal angle $\Phi$ can 
be easily integrated out since no polarization degrees of freedom are 
involved). Since $\Theta_{\pi^0}$ and $T_{\pi^0}$ are measured, the efficiency
can be corrected individually for each detected $\pi^0$ meson. This method 
generates a model independent result for the cross section as long as the
efficiency does not vanish for any kinematically possible combination of 
$\Theta_{\pi^0}$ and $T_{\pi^0}$. 
The simulated efficiency is plotted in the upper part of fig. \ref{fig:effi}. 
The plot demonstrates, that although the overall efficiency is not high, there 
is only a small region in the ($\Theta_{\pi^0}$,$T_{\pi^0}$) plane at kinetic 
energies larger than 200 MeV and forward angles smaller than 15$^o$ were 
no $\pi^0$ mesons can be detected. This is so, since in the 
energy range of interest almost always some fraction of the decay photon pairs
of $\pi^0$ mesons falls within the spectrometer acceptance.  Kinetic energies 
larger than 700 MeV 
are impossible for incident photon energies $\leq$800 MeV and the zero-efficiency 
region at backward angles also corresponds to a kinematically forbidden region.
As an example of a measured distribution the bottom part of fig. \ref{fig:effi}
shows the data for $^{40}$Ca. The small acceptance gap at forward angles was 
corrected with a phase space simulation. This is not model independent, but 
the largest corrections to the total cross sections are below 5\% so that 
possible systematic uncertainties arising from the acceptance gap are 
certainly below the 1\% level. 

The inclusive cross section $\sigma_{incl}$ determined this way can be written 
in the following form:
\begin{eqnarray}
\label{eq:incl}
\sigma_{incl} & = & \sigma_{\pi^0}+2\sigma_{2\pi^0}+\sigma_{\pi^0\pi^+}
              +\sigma_{\pi^0\pi^-}+\beta\sigma_{\eta}\\
& &\beta = 3b_{\eta\rightarrow 3\pi^0}+b_{\eta\rightarrow \pi^0\pi^+\pi^-}
\approx 1.2\nonumber 	      
\end{eqnarray}
where $\sigma_{x}$ are the partial cross sections labeled with the different
final states and $b_{i}$ are the branching ratios of the $\eta$ meson for 
the respective decay channels.
It was previously shown in \cite{Krusche_99} that this relation holds very 
accurately in case of the deuteron, where all partial cross sections are known.
This means in particular that final states with pion multiplicity larger than
two, which do not originate from decays of the $\eta$ meson, are negligible in
this energy range. Provided the partial cross sections for $\eta$ and double 
$\pi^0$ production are known, the part $\sigma_{nm}$ of the total 
photoabsorption cross section $\sigma_{abs}$ with at least one neutral 
pion or one $\eta$ meson in the final state follows from:
\begin{equation}
\label{eq:sigma_cor}
\sigma_{nm} = \sigma_{incl} - \sigma_{2\pi^0} - (\beta -1)\sigma_{\eta}\;\;. 
\end{equation}
The total photoabsorption cross section $\sigma_{abs}$ can then be split into:
\begin{equation}
\sigma_{abs}=\sigma_{nm}+\sigma_{cm}+\sigma_{r}
\end{equation}
where $\sigma_{cm}$ corresponds to reactions which have no neutral meson 
but at least one charged pion in the final state, and $\sigma_{r}$ corresponds 
to final states without any mesons. In case of the deuteron, $\sigma_{r}$ 
equals the deuteron photo-breakup.

\subsection{Exclusive reaction channels}
\label{ssec:ana_excl}
The decomposition of the inclusive process into exclusive reaction channels is
discussed in detail in \cite{Krusche_99} for the deuteron target, but it is
less straight forward for heavy nuclei. The larger Fermi momenta cause a 
stronger smearing of kinematical observables, and FSI affects the reaction 
products. The first step is the separation of quasifree single $\pi^0$ 
photoproduction. Since the solid angle coverage of the detector amounts only 
to $\approx$25\% of 4$\pi$ this must be done with cuts on the reaction 
kinematics. 
\begin{figure}[h]
\centerline{
\resizebox{0.48\textwidth}{!}{%
  \includegraphics{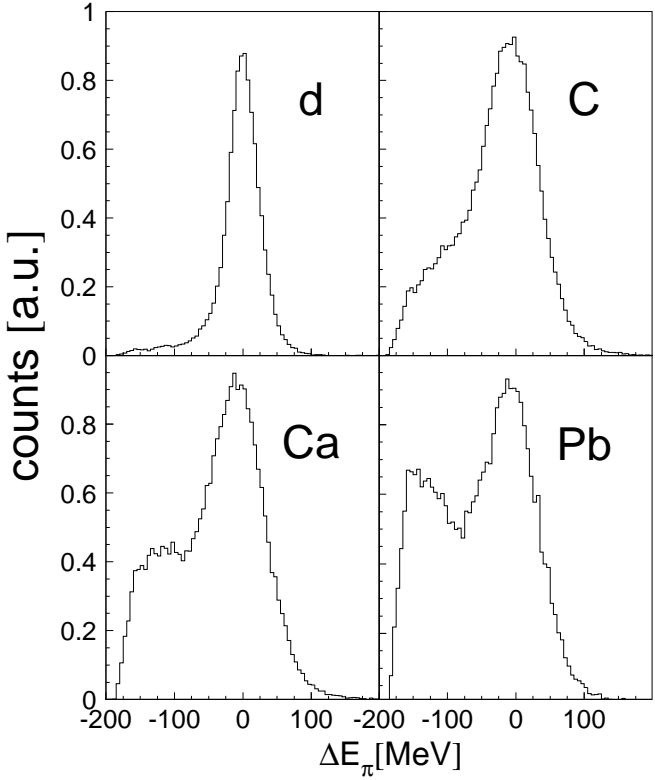}     
}}
\caption{Missing energy spectra for $\pi^0$ photoproduction for incident
photon energies 395 - 440 MeV calculated under the assumption of quasifree
kinematics. 
}
\label{fig:misse1}       
\end{figure}
Missing energy ($\Delta E_{\pi}$) spectra can be constructed for the kinetic 
energy $E^{\star}_{\pi}$ of the pion in the center-of-momentum (cm) frame of 
the incident photon and a stationary nucleon:
\begin{equation}
\Delta E_{\pi} = E^{\star}_{\pi}(\gamma_1\gamma_2) - E^{\star}_{\pi}(E_{b})\;\;.
\end{equation}
Here, $E^{\star}_{\pi}(E_{b})$ is the pion energy derived from the energy of 
the incident photon beam $E_{b}$ and $E^{\star}_{\pi}(\gamma_1\gamma_2)$ is 
reconstructed from the momenta of the pion decay photons.

The reaction kinematic is assumed to be quasifree, i.e. the pion is produced 
from one participant nucleon and the rest of the nucleus acts as spectator. 
The neglect of the Fermi momenta of the bound nucleons simply broadens the
missing energy peaks. At low incident photon energies a significant fraction 
of $\pi^0$ photoproduction stems from the coherent reaction, in which
the recoil is not taken by a nucleon but by the whole nucleus. 
A previously published analysis \cite{Krusche_02} has separated the coherent
parts with a missing energy analysis assuming coherent kinematics which gives 
much narrower peaks since no Fermi smearing is involved. In the present
analysis contributions from the coherent reaction fall within the broader peak 
of quasifree pion production.

\begin{figure}[h]
\centerline{
\resizebox{0.49\textwidth}{!}{%
  \includegraphics{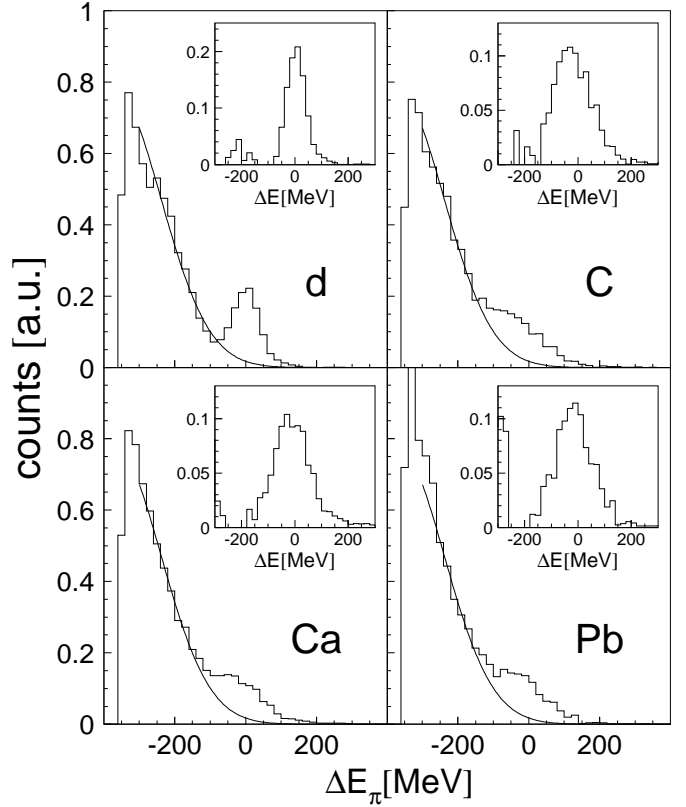}     
}}
\caption{Missing energy spectra for $\pi^0$ photoproduction for incident
photon energies of 790 MeV calculated under the assumption of quasifree
kinematics.
}
\label{fig:misse2}       
\end{figure}

An example of the missing energy spectra at medium incident photon
energy around 420 MeV is shown in fig. \ref{fig:misse1}. There is a striking
difference between the missing energy distribution from the deuteron and from 
heavy nuclei. At these energies, in case of the deuteron a peak with a 
small tail at negative values is visible. The tail is due to the onset of
double pion production which has, however, a relatively small cross section
in this energy region. On the other hand, the distribution for lead shows a
substantial non-quasifree background. At this photon energy the background
arises mainly from single pion production with significant final state 
interaction of the pion or from two-body contributions to the production 
process. It must be emphasized that the single $\pi^0$ production cross 
section $\sigma_{\pi^0}^{qf}$ obtained from this kinematical analysis is not 
identical with the cross section $\sigma_{\pi^0}$ in eq. \ref{eq:incl}. By 
definition $\sigma_{\pi^0}$ is the cross section for events with one and only 
one neutral pion in the final state, it does not matter if the quasifree 
energy relation is valid. However, single $\pi^0$ events, in which the 
$\pi^0$ suffers an FSI or was produced by a two-body interaction, will in 
general be excluded from $\sigma_{\pi^0}^{qf}$. Only for the deuteron, where 
FSI and two-body absorption do not significantly contribute, we have 
$\sigma_{\pi^0}^{qf}\approx \sigma_{\pi^0}$. 
At higher incident photon energies, much larger contributions from multiple 
pion final states are visible in the missing energy distributions. An example 
is shown in fig. \ref{fig:misse2}. This figure shows also the lineshape of the
missing energy peaks, which are broader for the heavy nuclei due to Fermi
motion. The shape of the background (solid curves in fig. \ref{fig:misse2})
was fitted to the deuteron data and then scaled for the nuclei. The
similarity of the background shapes for the deuteron and the heavy nuclei
is not surprising. The background for the deuteron and the nuclei
is dominated by the same source, namely double pion events where only one pion 
was detected. Folding the curve with a typical momentum distribution for 
nucleons bound in heavy nuclei has not much effect since already the slope 
for the deuteron is not steep. However, fitting and subtracting the background 
contributions would introduce systematic uncertainties which are difficult to
control for the comparison to model predictions. We have therefore extracted
the quasifree single $\pi^0$ production with a cut in the missing energy 
spectra from 0 - 200 MeV, assuming symmetric line shapes. Contributions 
from the tail of the background distribution were assumed to be negligible as
suggested by fig. \ref{fig:misse2}. The same analysis was applied to the 
results of the BUU calculations so that systematic effects in the comparison
between data and model results are minimized. 

The analysis of double $\pi^0$ photoproduction from the proton and the deuteron
is discussed in detail in \cite{Krusche_99,Haerter_97,Kleber_00,Wolf_00}.
At incident photon energies below the $\eta$ production threshold the cross
sections of all other reactions with at least three photons in the final state
are negligible compared to double $\pi^0$ production. Therefore, an
analysis of all events with at least three identified photons is sufficient for
the double $\pi^0$ channel. Background may only arise from particles which are
misidentified as photons or from false photon pairs which result from the
split-off of an electromagnetic shower into two geometrically unrelated
components. Particles were completely suppressed with the veto detectors, the 
pulse shape analysis and the time-of-flight analysis. 
Subsequently, a standard invariant mass analysis was used to assign two photons
(labeled $\gamma_{1}$, $\gamma_{2}$) to the decay of one $\pi^0$. The third
photon ($\gamma_{3}$) was then a candidate for a second $\pi^0$ decay with
one undetected photon. Among the latter, split-off photons were suppressed
with cuts on the relative timing ($\Delta t_{\gamma\gamma}\leq$1 ns for all
photon combinations), the minimum energy of the third photon 
($E_{\gamma_3}\geq$45 MeV), and the minimum opening angle between the third
photon and the other two photons ($\Theta_{\gamma_1\gamma_3}$, 
$\Theta_{\gamma_2\gamma_3}\geq$ 20$^o$). 

\begin{figure}[th]
\centerline{
\resizebox{0.5\textwidth}{!}{%
  \includegraphics{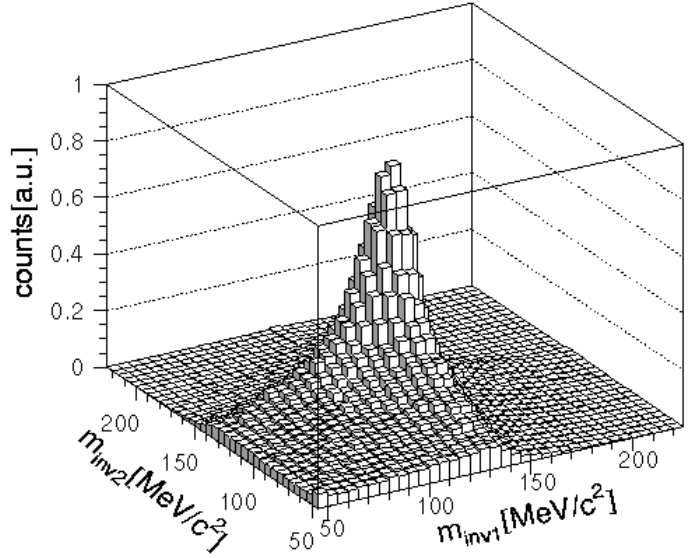}   
}}
\caption{Invariant mass of photon pair 1 versus invariant mass of photon pair
2 for events with four detected photons.
}
\label{fig:minv}       
\end{figure}

The above analysis was only used for the data measured with the first 
setup where the small solid angle coverage resulted in a very low detection 
efficiency for events with four photons. A much more restrictive analysis 
is used for events with four detected photons which can be combined into two 
pairs with the invariant mass of the pion. A two-dimensional representation
of the invariant masses of the photon pairs is shown in fig. \ref{fig:minv}.
All possible combinations of the four photons into two pions are 
tested and the `best' combination is chosen via a $\chi^2$ analysis. Only
these `best' combinations are shown in the figure. After a cut on the 
invariant mass peak a missing mass analysis as discussed in \cite{Krusche_99} 
serves as a kinematical constraint. 
The mass of the missing particle (the participant nucleon) is calculated 
from the momenta $\vec{P_{\gamma_i}}$ and the energies $E_{\gamma_i}$ 
of the pion decay photons and the beam energy $E_b$ and 
compared to the nucleon mass $m_N$. Fermi motion is again neglected.
The missing mass $\Delta M$ is given by: 
\begin{equation}
\Delta M = \sqrt{(E_b+m_N-\sum_{i=1}^{4}{E_{\gamma_i}})^2
-(\vec{P_b}-\sum_{i=1}^{4}{\vec{P_{\gamma_i}}})^2}- m_N.
\end{equation}

\begin{figure}[th]
\centerline{
\resizebox{0.47\textwidth}{!}{%
  \includegraphics{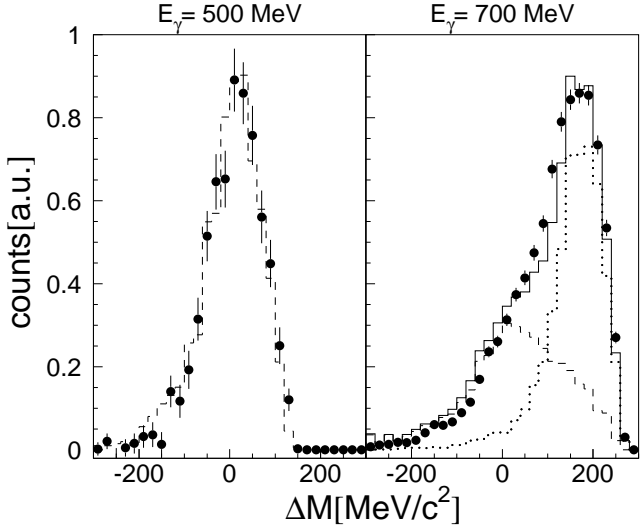}     
}}
\caption{Missing mass spectra for double $\pi^0$ photoproduction from
Ca. Left side: incident photon energy of 500 MeV (below $\eta$ threshold),
Right side: incident photon energy 700 MeV (above $\eta$ threshold).
Closed circles: experiment, dashed histograms simulation of 
$2\pi^0$ photoproduction,
dotted histogram: simulated contribution of $\eta$ photoproduction,
solid histogram: sum of both.
}
\label{fig:mismas}       
\end{figure}

Missing mass spectra for calcium below and above the $\eta$ production
threshold are shown in fig. \ref{fig:mismas}. 
At incident photon energies below the $\eta$ production threshold only a broad
peak around zero is visible which agrees with the result of a Monte 
Carlo simulation taking into account Fermi motion, a simple approximation
for the final state interaction, and the instrumental acceptance and detection
efficiency. At higher incident photon energies background
from the $\eta\rightarrow 3\pi^0$ decay appears. 
In contrast to the deuteron case \cite{Krusche_99,Kleber_00}, the broadening 
of the peak due to Fermi motion and FSI is so large that no clear separation 
from the $\eta$ background is possible.
However, since the cross section for $\eta$ photoproduction 
from nuclei is known from the analysis of the $2\gamma$ decay channel 
of the $\eta$ meson it can be subtracted. Simulations of double
$\pi^0$ production and the $\eta$ background are compared to the data in
fig. \ref{fig:mismas}. The shape of the sum of the two contributions again 
agrees with the data.

An analogous analysis was done for the $\pi^0\pi^{\pm}$ reaction. The
relative contribution of the $\eta$ background is much smaller in this case.
For the free nucleon 
$\sigma(\gamma N\rightarrow \pi^0\pi^{\pm})$ is almost six times larger than
$\sigma(\gamma N\rightarrow \pi^0\pi^0)$,
while the branching ratio for $\eta\rightarrow\pi^+\pi^-\pi^0$ is a factor
of 1.4 smaller than for $\eta\rightarrow 3\pi^0$. 

The detection efficiency for double $\pi^0$ events cannot 
be constructed in the model independent way used for inclusive and single 
$\pi^0$ photoproduction since the angles and energies of the pions are  
correlated. The case of $\pi^0\pi^{\pm}$ photoproduction is even more
difficult since certain regions of the polar angle of the charged pions
are not within the acceptance of the detector (spaces between blocks).  
Therefore a phase space simulation of the reactions, including the effects 
of Fermi motion and a simple approximation of meson final state interaction
was used. For the latter only re-scattering of the pions was considered
and the strength of this FSI was varied until all angular, energy and 
missing mass distributions of the input to the simulations where consistent 
with the data (see e.g. fig. \ref{fig:mismas}). From the comparison of 
different simulations and the good agreement between the analysis of the
three-photon events from the first measurement and the four-photon events
from the second measurement of double $\pi^0$ production 
(see fig. \ref{fig:2pi0}) we estimate a systematic uncertainty of 5\%.

\section{The BUU-model}
\label{sec:BUU}
In this section we discuss the main ingredients of the BUU transport model.
For the present calculations we use the version discussed in detail in
\cite{Effenberger_99,Lehr_00}.

The model is based on the BUU equation which describes the space-time 
evolution ($\vec{r}$: space coordinate, $\vec{p}$: momentum) of the spectral 
phase-space density $F_i$ of an ensemble of interacting particles of 
type $i$ = $N$, $P_{33}(1232)$, $\pi$, $\eta$,... with mass $\mu$:  
\begin{eqnarray}
  \left({\partial\over\partial t}+\vec{\nabla}_p H\cdot\vec{\nabla}_r
  -\vec{\nabla}_r H\cdot\vec{\nabla}_p\right)F_i(\vec r,\vec p,\mu;t)=\nonumber
\\  
= I_{coll}[F_N,F_\pi,F_{P_{33}(1232)},F_\eta,...]
\end{eqnarray}
The left-hand side -- the Vlasov term -- describes the propagation 
of the particles under the influence of a Hamilton function $H$, given
by the expression
\begin{equation}
  H=\sqrt{(\mu+S)^2+p^2},
\end{equation}
which in the case of baryons contains an effective scalar potential $S$
\cite{Lehr_00}. 
The right-hand side of the BUU equation (called the collision integral) 
consists of a gain and a loss term for the phase space density $F_i$ at the 
different space-time points, accounting for interactions between the particles
beyond the mean-field potential.
The collision integral contains collision rates for the different reaction 
types such as baryon-baryon and baryon-meson collisions, resonance formation 
and decay. They include cross sections for these processes and 
Pauli blocking factors for outgoing fermions. Due to the reactions of 
different particle types we end up with a set of coupled BUU equations,
which is solved by a test particle ansatz. This converts the continuous
phase space density into an ensemble of discrete test particles propagating
according to Hamilton's equations of motion.

Before the actual reaction, the nucleon test particles are initialized in the
nucleus according to a Woods-Saxon density function in coordinate space. In 
momentum space, we make use of the local density approximation, distributing 
the nucleons homogeneously in the Fermi sphere corresponding to the local
density. For the baryons, the mean-field potential $S$ is determined as 
described in \cite{Lehr_00}. In this work, we focus on a momentum dependent 
potential corresponding to an equation of state with medium compressibility.

The photon-nucleus reaction is modeled in terms of the absorption
of the photon on a single nucleon, leading to final states P$_{33}$(1232),
D$_{13}$(1520), S$_{11}$(1535), F$_{15}$(1680), $N\pi$, $N\pi\pi$,
$NV$ ($V$=$\rho$, $\omega$, $\phi$), $K\Lambda$, $K\Sigma$ and $K\bar K N$.
These entry states are selected with probabilities corresponding to the 
different reaction cross sections. They are then subjected to the final state
interactions, which are described by the set of BUU equations. The cross 
sections for the $\gamma A$ reaction are then determined by averaging 
over an ensemble of such elementary reactions as outlined in 
\cite{Effenberger_97}.

Besides Fermi motion, binding effects and Pauli blocking we also account
for collisional broadening of the most important resonances P$_{33}$(1232),
D$_{13}$(1535) and S$_{11}$(1535). For the in-medium width of the P$_{33}$
we either use the results of Oset et al. \cite{Oset_87} or the phenomenological 
spreading potential of Hirata et al. \cite{Hirata_79}, both giving roughly 
$\Gamma_{coll}=80$ MeV at saturation density $\rho_o$. The latter is 
interpreted  as fully absorptive, whereas the absorption part in 
\cite{Oset_87} is only about 40 MeV at $\rho_o$. For the D$_{13}$(1520) and 
the S$_{11}$(1535), we make use of the results of the self-consistent 
resonance-hole model in \cite{Post_03}. In particular, for the D$_{13}$ a 
total collision width of $200\rho/\rho_o$ MeV at density $\rho$ is used, 
while the S$_{11}$ collision width is rather small and amounts only to 
$30\rho/\rho_o$ MeV. In both cases, one quarter of the total width is used 
for the absorptive channels.

\section{Results and discussion}
\label{sec:4}

\subsection{Inclusive $\pi^0$-photoproduction}
\begin{figure}[hb]
\resizebox{0.49\textwidth}{!}{%
  \includegraphics{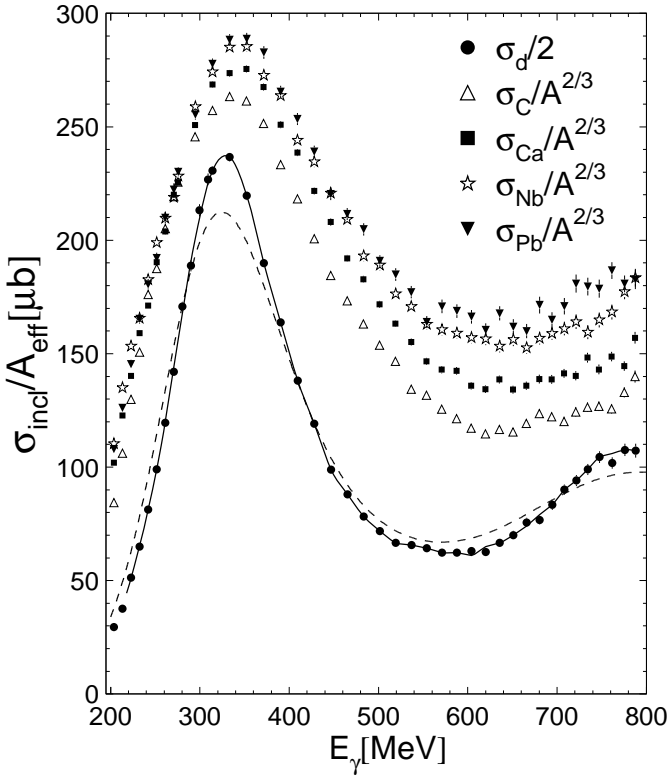}     
}
\caption{Scaled total cross section $\sigma_{incl}$ of inclusive 
$\pi^0$ photoproduction from
deuterium, carbon, calcium, niobium, and lead  versus incident photon energy
(scale factor 1/2 for the deuteron, $A^{2/3}$ for all other nuclei). 
The solid line is only to guide the eye, the dashed 
line corresponds to the deuteron data folded with the momentum 
distribution of nucleons in calcium.
}
\label{fig:totincl}       
\end{figure}

The total inclusive $\pi^0$ cross section $\sigma_{incl}$ for the heavy 
nuclei is compared in fig. \ref{fig:totincl} to the deuteron cross section.
The cross sections are scaled by $A_{eff}$ which throughout the paper is defined
as $A_{eff}$=2 for the deuteron and $A_{eff}=A^{2/3}$ for the other nuclei.
In this way, the average nucleon cross section is compared to the nuclear
cross sections scaled in proportion to the nuclear surfaces. 
The shape of the total cross sections for the heavy nuclei is similar, but it
differs from the deuteron. 
The peak in the $\Delta$ resonance region is broader for the heavier nuclei 
and shifted to a higher energy. Furthermore, the valley 
between the first and second resonance region is much more pronounced for the 
deuteron. The different shape is not explained by the larger Fermi momenta of 
the nucleons bound in the heavier nuclei. This is demonstrated with
the dashed curve in fig. \ref{fig:totincl} which represents the deuteron cross 
section folded with the nucleon momentum distribution of Ca. The latter was 
parameterized as \cite{Cassing_94}:
\begin{eqnarray}
P_F(p)dp & = &
k_c\cdot p^2\cdot\left(e^{-p^2/a}+ce^{-p/b}\right)dp\\
k_c & = & 4\left(\pi^{1/2} a^{3/2}+8cb^3\right)^{-1}\nonumber
\end{eqnarray}
with $a=0.42$ fm$^2$, $b=0.23$ fm, and c=0.04. The folding was done as 
discussed in \cite{Krusche_95b}. Here, it was neglected that the deuteron 
excitation function already incorporates the (much smaller) Fermi smearing 
effects from the deuteron wave function which in principle would have to be 
de-folded before folding with the Ca distribution.  

It is tempting to argue that the shift of 
the peak in the $\Delta$ region by $\approx$20 MeV is related to an in-medium 
modification of the $\Delta$ resonance due to the $\Delta$ - nucleus 
interaction. Analyses of coherent $\pi^0$ photoproduction gave indications of a 
$\Delta$ self-energy corresponding to a similar peak shift, almost independent 
of the nuclear mass \cite{Rambo_99,Drechsel_99,Krusche_02}. However, the 
interpretation is not so simple. The shape of the low energy side of the peak 
reflects the superposition of contributions from coherent and quasifree single 
$\pi^0$ photoproduction which have different energy dependencies. 
Furthermore, as discussed below, contributions from multi-body absorption 
effects of the incident photons are important in the high energy tail
of the $\Delta$. Altogether, position and shape of the $\Delta$ peak are 
influenced by several effects of different nature and not simply connected to 
the in-medium properties of the $\Delta$.

\begin{figure}[t]
\resizebox{0.48\textwidth}{!}{%
  \includegraphics{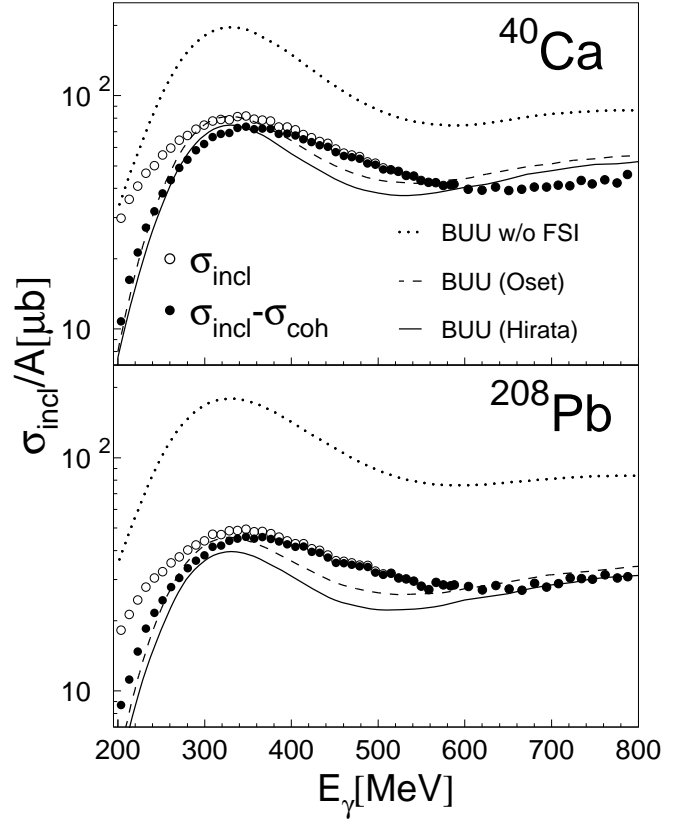}     
}
\caption{Comparison of the inclusive cross section $\sigma_{incl}$ to BUU 
model calculations 
for calcium and lead. Open symbols: inclusive reaction, filled symbols:
difference of inclusive and coherent reaction. Curves: BUU without FSI 
(dotted), BUU with FSI, $\Delta$ in-medium width from \cite{Oset_87} (dashed) 
or \cite{Hirata_79} (solid).}
\label{fig:lehr_1}       
\end{figure}

The data for calcium and lead are compared to the BUU calculations in fig.
\ref{fig:lehr_1}. Since by construction the BUU model accounts only for
the incoherent part of photoproduction, the data are also shown after 
subtraction of the contribution from coherent single $\pi^0$ production
taken from \cite{Krusche_02}. 
The figure demonstrates the strong absorptive character of the FSI. The
calculation without FSI overestimates the lead data almost by a factor of
four, but the inclusion of FSI produces a reasonable description of
the data. The two models for the P$_{33}$ in-medium widths lead to similar
results. The inclusion of the spreading potential gives a larger
suppression of the cross section. The broad structure of the first
resonance peak cannot be reproduced by the values used for the P$_{33}$
collision width but an increase much beyond 100 MeV at density $\rho_o$ would
lead to an even stronger suppression of the cross section and is not
justified by theoretical models. However, it is known \cite{Carrasco_92} 
that two-body absorption mechanisms like $\gamma$NN$\rightarrow N\Delta$ 
are non-negligible in this energy range. Such contributions are not included 
in the BUU model.

\begin{figure}[thh]
\resizebox{0.5\textwidth}{!}{%
  \includegraphics{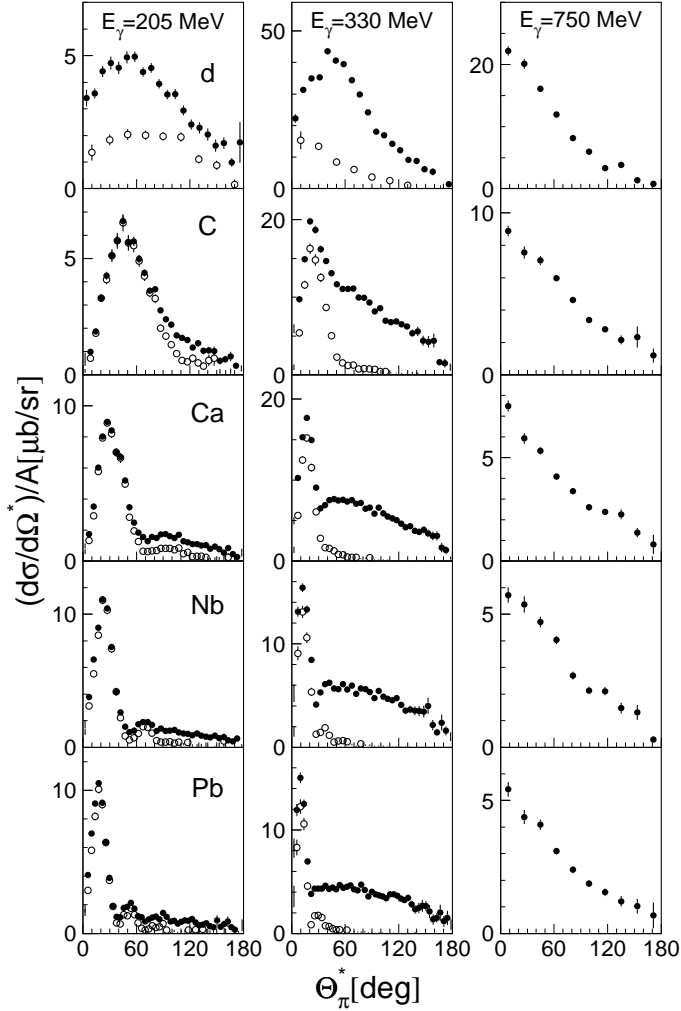}     
}
\caption{Angular distributions in the photon - nucleus cm frame scaled by the
mass number. Full symbols: inclusive cross section this work, open symbols: 
coherent $\pi^0$ production \cite{Krusche_02}.
}
\label{fig:ang1}       
\end{figure}

\begin{figure}[hhh]
\resizebox{0.48\textwidth}{!}{%
  \includegraphics{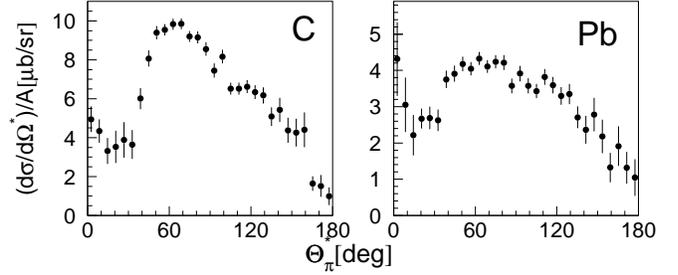}     
}
\caption{Angular distributions without coherent $\pi^0$ photoproduction
in the photon - nucleus cm frame scaled by the mass number for incident photon
energies from 325 - 335 MeV. 
}
\label{fig:angdis2}       
\end{figure}

Typical angular distributions are summarized in 
fig. \ref{fig:ang1}. The contribution from coherent single $\pi^0$
photoproduction is clearly visible for incident photon energies up to the 
$\Delta$ resonance region. It disappears at higher photon energies where the 
distributions are similar for all nuclei. The forward peaking at high photon
energies is a consequence of the frame choice (photon - nucleus cm system).
Due to the lack of the coherent part, the model calculations cannot reproduce
the low energy angular distributions. Subtraction of the coherent part would 
not help, since as discussed for the deuteron in 
\cite{Krusche_99,Siodlaczek_01} the coherent and quasifree parts are somewhat 
interconnected via nucleon final state interaction. The effect is also 
observed for the heavy nuclei.
As an example the angular distributions of the quasifree part around 
incident photon energies of 330 MeV are shown in fig. \ref{fig:angdis2}. 
The angular distributions are significantly depleted at forward angles where 
some strength is re-dis\-tri\-buted to the coherent part. 

\begin{figure}[h]
\resizebox{0.5\textwidth}{!}{%
  \includegraphics{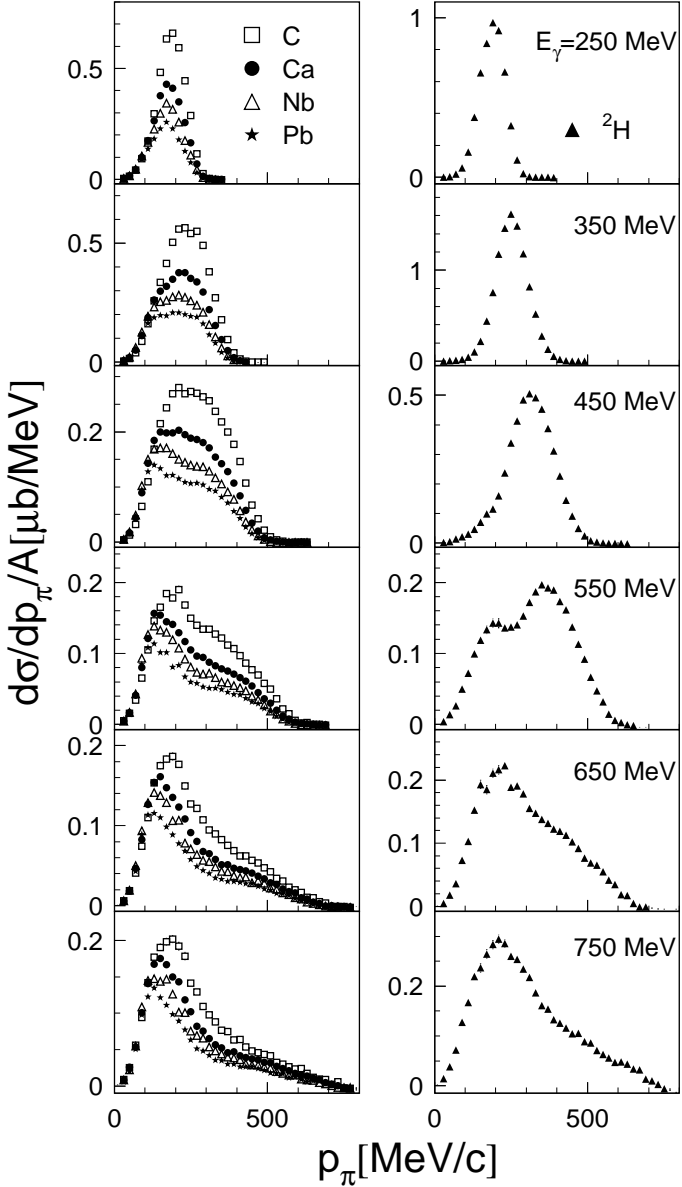}     
}
\caption{Distributions of the pion lab momenta for different intervals of
incident photon energy. Left hand side: heavy nuclei, right hand side:
deuterium.
}
\label{fig:inclm}       
\end{figure}

\begin{figure}[t]
\resizebox{0.48\textwidth}{!}{%
  \includegraphics{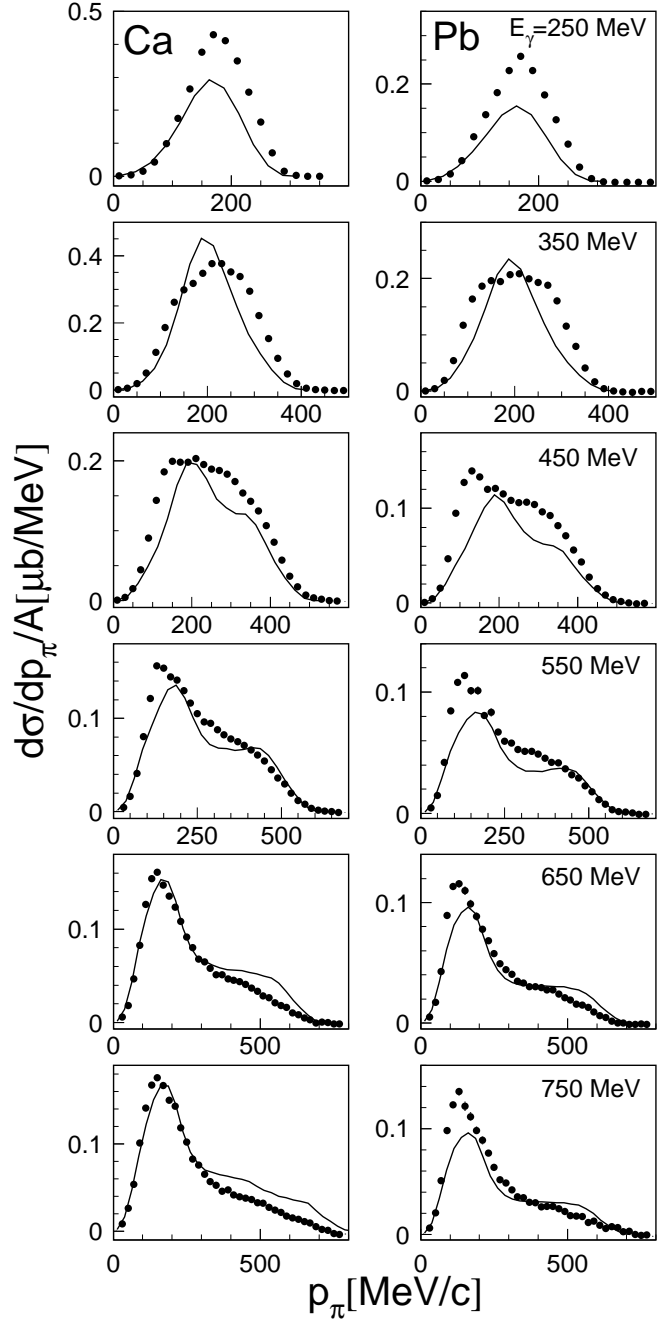}     
}
\caption{Distributions of the pion lab momenta for calcium (left hand side)
and lead (right hand side) compared to BUU calculations. Data and model
results are averaged over 100 MeV wide bins of incident photon energies,
the central values are given in the plot. The calculation used 
the spreading potential of Hirata et al. 
}
\label{fig:lehr_mom}       
\end{figure}

The momentum distributions of the pions are summarized in fig. \ref{fig:inclm}.
These spectra show the influence of FSI effects. Apart from two body
effects in the absorption of the photons, the distributions from the deuteron
can be regarded as source spectra of the pions, which are then modified for
the heavier nuclei by FSI. At the lowest incident photon energies spectra from 
the deuteron and the heavy nuclei are similar. Up to energies of 500 MeV the 
distributions measured on the deuteron simply shift to larger momenta 
corresponding to the quasifree kinematics of single $\pi^0$ production. 
However, the nuclei in this energy region show already a re-distribution of 
the pion momenta to smaller values. At small momenta the nuclear data scale 
with the mass number $A$ which indicates a large mean free path of the pions 
in this range. At even higher incident photon energies the pion spectra from 
the deuteron develop a double bump structure. The additional peak at lower 
momenta comes from double pion production reactions. At the highest energies 
this contribution becomes dominant. The contribution from single pion 
production for the heavier targets is suppressed due to FSI effects even 
at lower incident photon energies. The data for calcium and lead are compared 
to the BUU results obtained with the spreading potential \cite{Hirata_79} in 
fig. \ref{fig:lehr_mom}. The discrepancy at the lowest incident photon 
energies is mainly due to coherent $\pi^0$ photoproduction which contributes 
to the data but is not included in the BUU calculations. The general trend of 
the FSI effects, i.e. the shift of strength to small pion momenta, is 
reproduced by the model.   

\begin{figure}[t]
\resizebox{0.48\textwidth}{!}{%
  \includegraphics{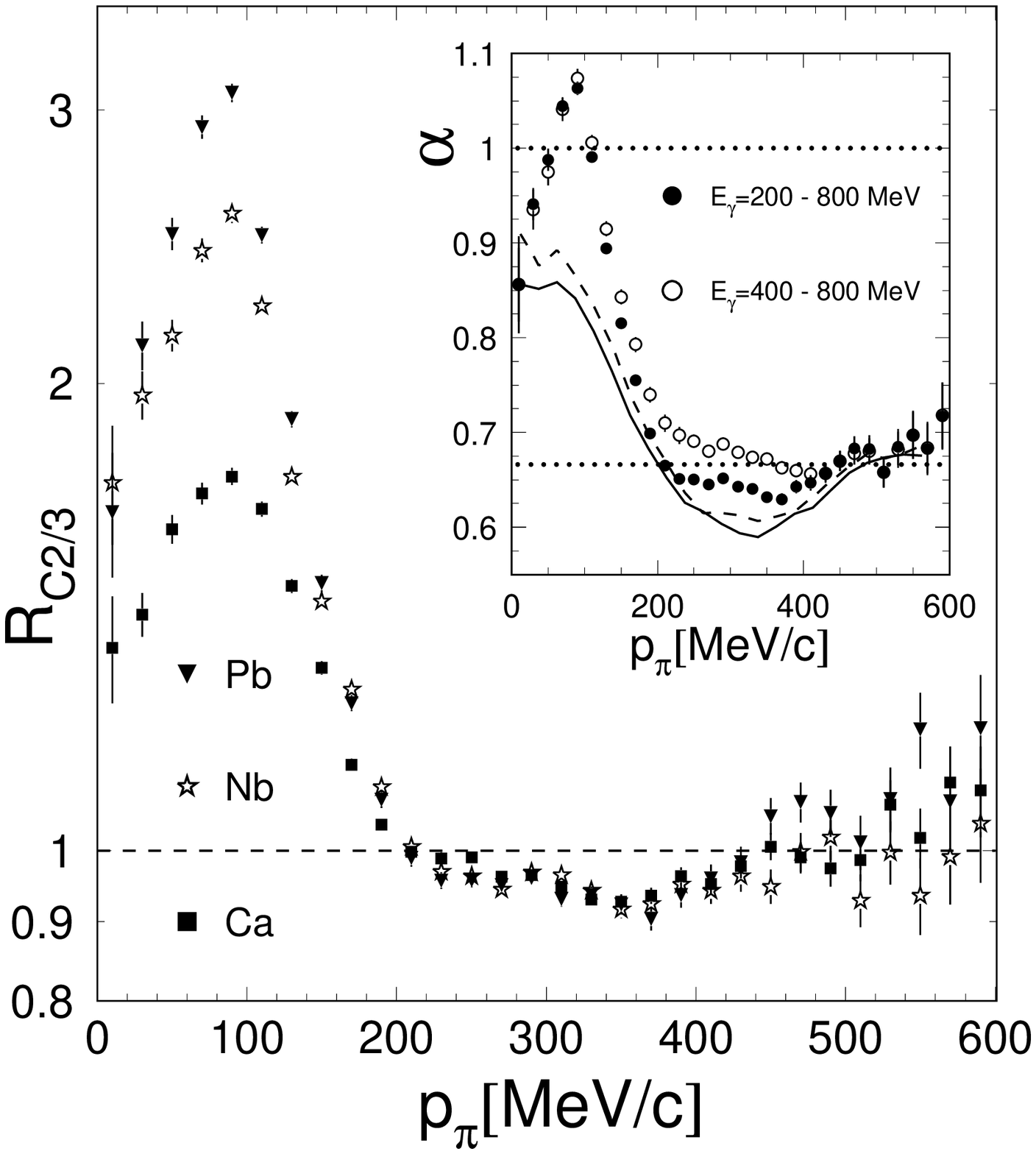}     
}
\caption{Scaling of the momentum dependence of the differential cross section 
with the nuclear mass number A. The main plot shows the ratio $R_{C2/3}$ (see eq. \ref{eq:rc23}) 
for Ca, Nb and Pb. The insert shows the fitted coefficient $\alpha$
(see eq. \ref{eq:alpha}) averaged over two ranges of incident photon energy.
Curves: BUU results (solid: spreading potential \cite{Hirata_79}, dashed: Oset 
parameterization \cite{Oset_87}). Both averaged over 400 - 800 MeV photon 
energy.
}
\label{fig:incls}       
\end{figure}

The FSI effects are closely related to the absorption properties of nuclear
matter for pions, that is the mean free path of pions in nuclear matter.
The momentum dependence of this effect is analyzed in fig. \ref{fig:incls}.
The main plot shows the ratio of the cross sections from the three heaviest
nuclei with respect to carbon, normalized to $A^{2/3}$. It is defined via:
\begin{equation}
\label{eq:rc23}
R_{C2/3}=\frac{\sigma (A)/A^{2/3}}{\sigma (^{12}C)/12^{2/3}}\;\;.
\end{equation}  
The insert shows the scaling exponent $\alpha$ obtained from fits of the 
mass dependence of the data with the ansatz:
\begin{equation}
\label{eq:alpha}
\sigma (A)\propto A^{\alpha}\;\;.
\end{equation}
The differential cross sections are averaged over the entire range
of incident photon energies from 200 - 800 MeV. The scaling coefficient
is also shown for incident photon energies above 400 MeV where contributions
from coherent $\pi^0$ photoproduction are negligible. Coherent contributions
could obscure the scaling, since their production (before FSI) is not simply
proportional to the mass number A. However, the plot shows that their
influence is small. Both figures demonstrate 
qualitatively the same behavior. For pion momenta larger than roughly 200 MeV/c
the cross sections scale like $A^{2/3}$ which indicates strong absorption
and small mean free paths of the pions. At smaller pion momenta the nuclei
become more transparent for pions and the cross sections scale with the 
volume. This is expected since pions with a momentum of 227 MeV/c can excite 
a nucleon to the nominal mass of the $\Delta$. Therefore, at momenta larger 
than $\approx$200 MeV/c
pions are very likely re-absorbed via $\pi N\rightarrow\Delta$. This explains
also qualitatively the shift of strength to small pion momenta in fig. 
\ref{fig:inclm}. Only those pions which are produced with small momenta or
acquire small momenta after absorption and re-emission processes are likely
to escape from the nucleus. Contributions from larger pion momenta
stem mainly from the nuclear surface region. These momentum distributions are
therefore very sensitive to the FSI effects and thus a valuable testing ground
for model predictions. The BUU calculations show qualitatively similar
behavior, however they do not reproduce the strong peak at pion momenta
around 100 MeV/c. This could indicate that the mean free path for low pion 
momentum is underestimated. However, one should be aware that
the coefficient $\alpha$ does not only depend on the mean free path of the 
pions. It can be also influenced to some extent by the re-distribution of the 
pion momenta due to the absorption and re-emission processes.

\begin{figure}[h]
\resizebox{0.48\textwidth}{!}{%
  \includegraphics{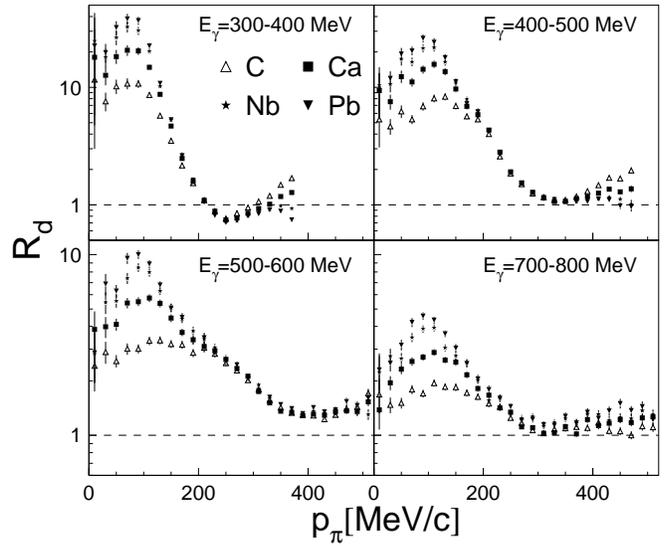}     
}
\caption{Ratio of cross section from heavy nuclei and deuterium. 
The ratio $R_d$ (see eq. \ref{eq:rd}) for carbon, calcium, niobium and lead
is plotted.
}
\label{fig:deura}       
\end{figure}
We have already noted in the discussion of the total inclusive cross section
that in case of the heavier nuclei the valley between the first and second 
resonance region is much shallower than for the deuteron. This could be due 
either to FSI effects or to the initial photo-pion production mechanism.  
The two possibilities can be tested by a comparison of the momentum spectra 
to the deuteron data.
For this purpose the ratio:
\begin{equation}
\label{eq:rd}
R_{d}=\frac{\sigma (A)/A^{2/3}}{\sigma (^2H)/2}
\end{equation}
which compares the nuclear data to the average nucleon cross section is plotted
in fig. \ref{fig:deura} as function of the pion momenta for different regions of
incident photon energies. At a given pion momentum the absorption probability
should be independent of the incident photon energy. For the lowest 
incident photon energies the ratio shows the expected behavior which reflects 
the momentum dependence of the pion mean free path discussed above: It is 
large for small pion momenta and then drops to unity around 
$p_{\pi}\approx$ 200 MeV/c. However, at incident photon energies between 
500 - 600 MeV the ratio is always larger than unity, in particular for momenta 
in the range between 200 to 350 MeV/c. This is an indication that
for the heavy nuclei in the intermediate energy range a certain fraction of 
pions is indeed produced via mechanisms which do not operate or are less
important for the deuteron. Possible candidates are contributions from 
multi-body absorption mechanisms like $\gamma NN\rightarrow N\Delta$ 
(see e.g. \cite{Carrasco_92}), in-medium effects of any of the involved 
partial reaction channels, or charge exchange reactions which effectively 
transfer strengths from final states like $\pi^+\pi^-$ to the inclusive 
$\pi^0$ channel. However, the latter are a less likely cause since the BUU 
calculations do not reproduce the additional strength at intermediate photon 
energies although they include the charge exchange reactions.

\subsection{Semi-exclusive reaction channels}
\begin{figure}[hth]
\centerline{\resizebox{0.49\textwidth}{!}{%
  \includegraphics{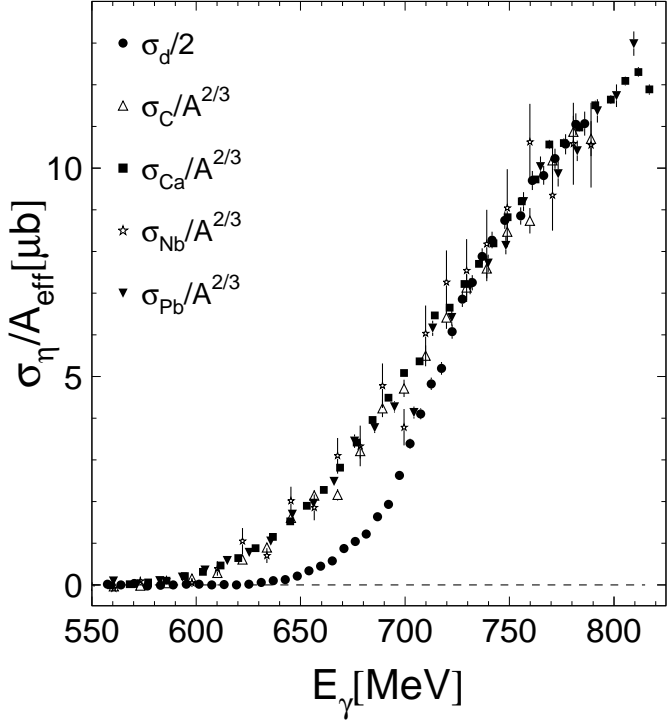}     
}}
\caption{Total cross section $\sigma_{\eta}$ of $\eta$ photoproduction from 
heavy nuclei. The data for carbon and niobium are taken from 
\cite{Roebig_96}, the data for calcium and lead are from the present
analysis of the measurements with the improved setup. The nuclear cross 
sections  are normalized to $A^{2/3}$ and compared to the average cross 
section from the nucleon ($\sigma_d /2$) \cite{Krusche_95b}. 
}
\label{fig:eta}       
\end{figure}

In the following we will decompose the inclusive cross section into partial 
reaction channels. The photoproduction of $\eta$ mesons contributes to the 
inclusive $\pi^0$ cross section via the $\eta\rightarrow 3\pi^0$ and 
$\eta\rightarrow \pi^0\pi^+\pi^-$ decays. Since the $\eta$ is long lived, 
it always decays outside the nuclei, so that these pions do not undergo FSI 
effects. The contribution to pion production is obtained directly from the 
measured $2\gamma$ decay channel of the $\eta$ mesons and the 
respective branching ratios. The scaled total cross sections $\sigma_{\eta}$ 
for $\eta$ photoproduction from nuclei are compared in fig. \ref{fig:eta} 
to the deuteron data \cite{Krusche_95b}. In case of carbon and niobium
previous data from \cite{Roebig_96} are used. 
For calcium and lead results with much improved statistical quality, obtained 
with the improved setup, are shown.
The nuclear data follow an $A^{2/3}$ scaling and agree above the production 
threshold on the free nucleon ($E_{thr}$=706 MeV) with the average nucleon 
cross section. In the immediate threshold region the reaction is only possible 
for large momenta of the bound nucleons 
anti-parallel to the incident photon momentum. The reaction on the deuteron is 
therefore suppressed in the threshold region since the Fermi momenta are much 
smaller than for the heavier nuclei.   
\begin{figure}[h]
\centerline{\resizebox{0.49\textwidth}{!}{%
  \includegraphics{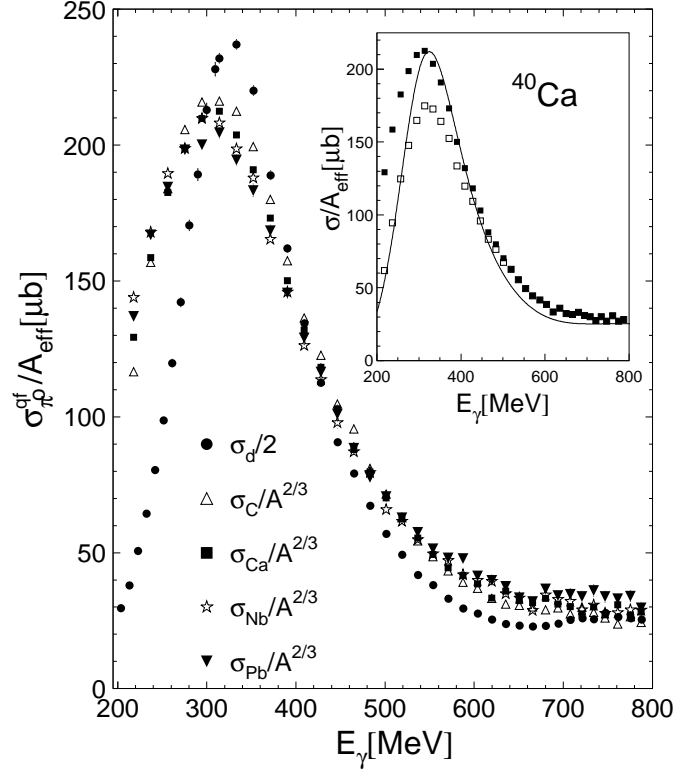}     
}}
\caption{Total cross section $\sigma_{\pi^0}^{qf}$ of single quasifree $\pi^0$ 
photoproduction from heavy nuclei. The nuclear cross sections are normalized 
to $A^{2/3}$ and compared to the average cross section from the nucleon 
($\sigma_d /2$). The insert compares for $^{40}$Ca $\sigma_{\pi^0}^{qf}$
(full points) to the cross section with subtracted coherent part. 
The curve corresponds to the deuteron cross section ($\sigma_d/2$) folded with 
the momentum distribution of nucleons in calcium. 
}
\label{fig:sipi}       
\end{figure}

Quasifree single $\pi^0$ photoproduction was extracted with the kinematical 
cuts discussed in sec. \ref{ssec:ana_excl}. Results for the second resonance 
region were already published in \cite{Krusche_01}. Again the nuclear cross 
sections $\sigma_{\pi^0}^{qf}$ scale like $A^{2/3}$ (see fig. \ref{fig:sipi}) 
and are similar to the average nucleon cross section. The difference between 
the scaled nuclear and deuteron cross sections at low incident photon energies 
is related to coherent $\pi^0$ photoproduction which is more important for 
the heavier nuclei. Note that in this channel in contrast to the inclusive 
reaction the position of the $\Delta$ peak is shifted to a lower incident 
photon energy for the heavier nuclei.

\begin{figure}[th]
\centerline{\resizebox{0.48\textwidth}{!}{%
  \includegraphics{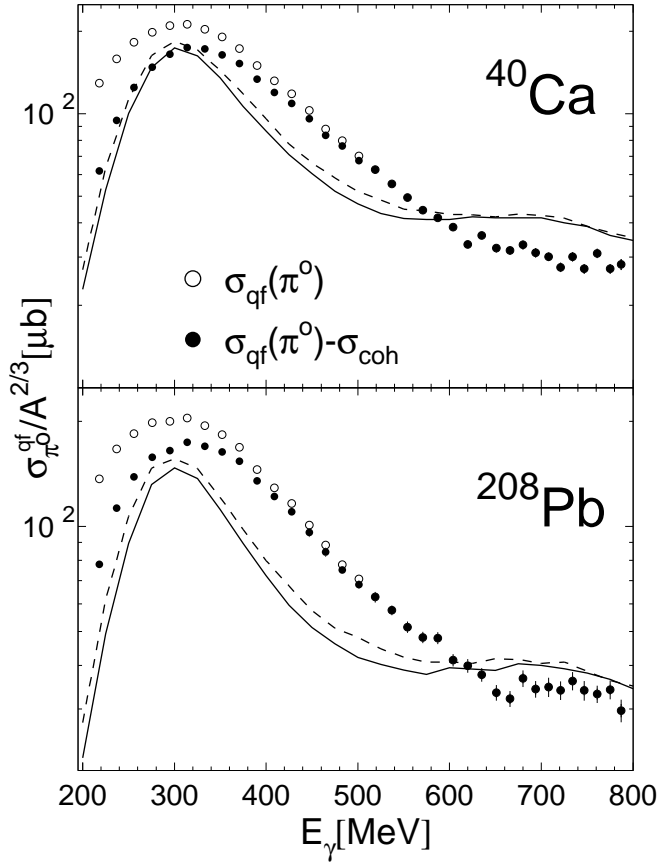}     
}}
\caption{Comparison of the single $\pi^0$ cross section to BUU model 
calculations for calcium and lead. Open symbols: single $\pi^0$ production, 
filled symbols: coherent part subtracted. Curves: BUU, 
$\Delta$ in-medium width from Oset parameterization (dashed) \cite{Oset_87}
or Hirata (solid) \cite{Hirata_79}. 
}
\label{fig:l_sipi}       
\end{figure}

In fig. \ref{fig:l_sipi}, the data for calcium and lead are compared to the 
results of the BUU model. Again, the data are also plotted after subtraction of
contributions from coherent $\pi^0$ production to facilitate a better 
comparison with the model. The calculations include the same missing energy 
cuts as the experiment. The curves show a similar behavior as in the case of 
the inclusive cross section in fig. \ref{fig:lehr_1}. In particular, the 
discrepancy at intermediate photon energies also shows up 
here. This suggests that two-body photon absorption mechanisms also 
contribute to this channel. In the second resonance region, the calculations 
overestimate the data. However, this does not necessarily mean that this is 
solely due to an overestimate of the D$_{13}$ contribution which in the 
calculations only amounts to roughly 25~\% of the total cross section for 
$E_\gamma=700$ MeV.

\begin{figure}[t]
\center{\resizebox{0.47\textwidth}{!}{%
  \includegraphics{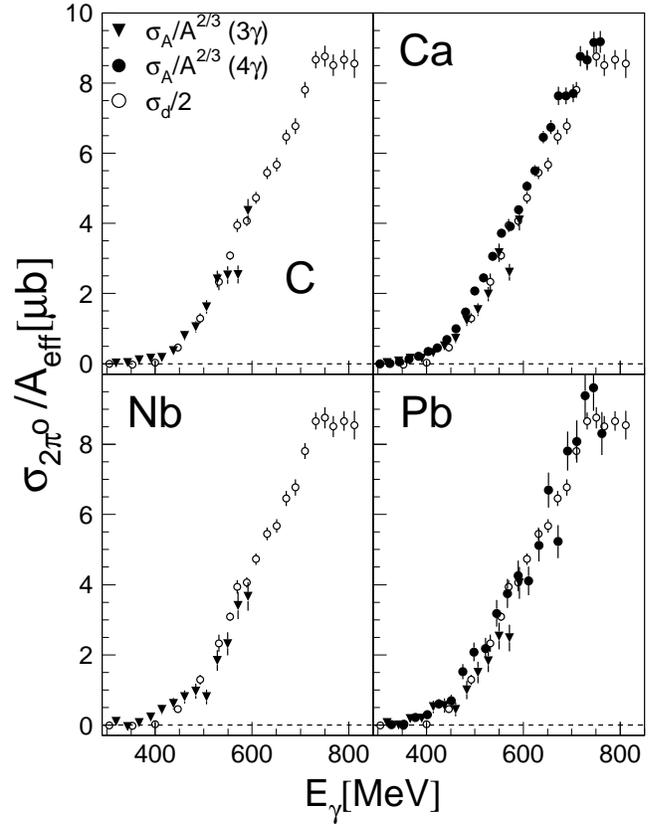}     
}}
\caption{Total cross section $\sigma_{2\pi^0}$ of double $\pi^0$ 
photoproduction normalized to $A^{2/3}$ and compared to the average nucleon 
cross section ($\sigma_d /2$). 
}
\label{fig:2pi0}       
\end{figure}

\begin{figure}[t]
\center{\resizebox{0.49\textwidth}{!}{%
  \includegraphics{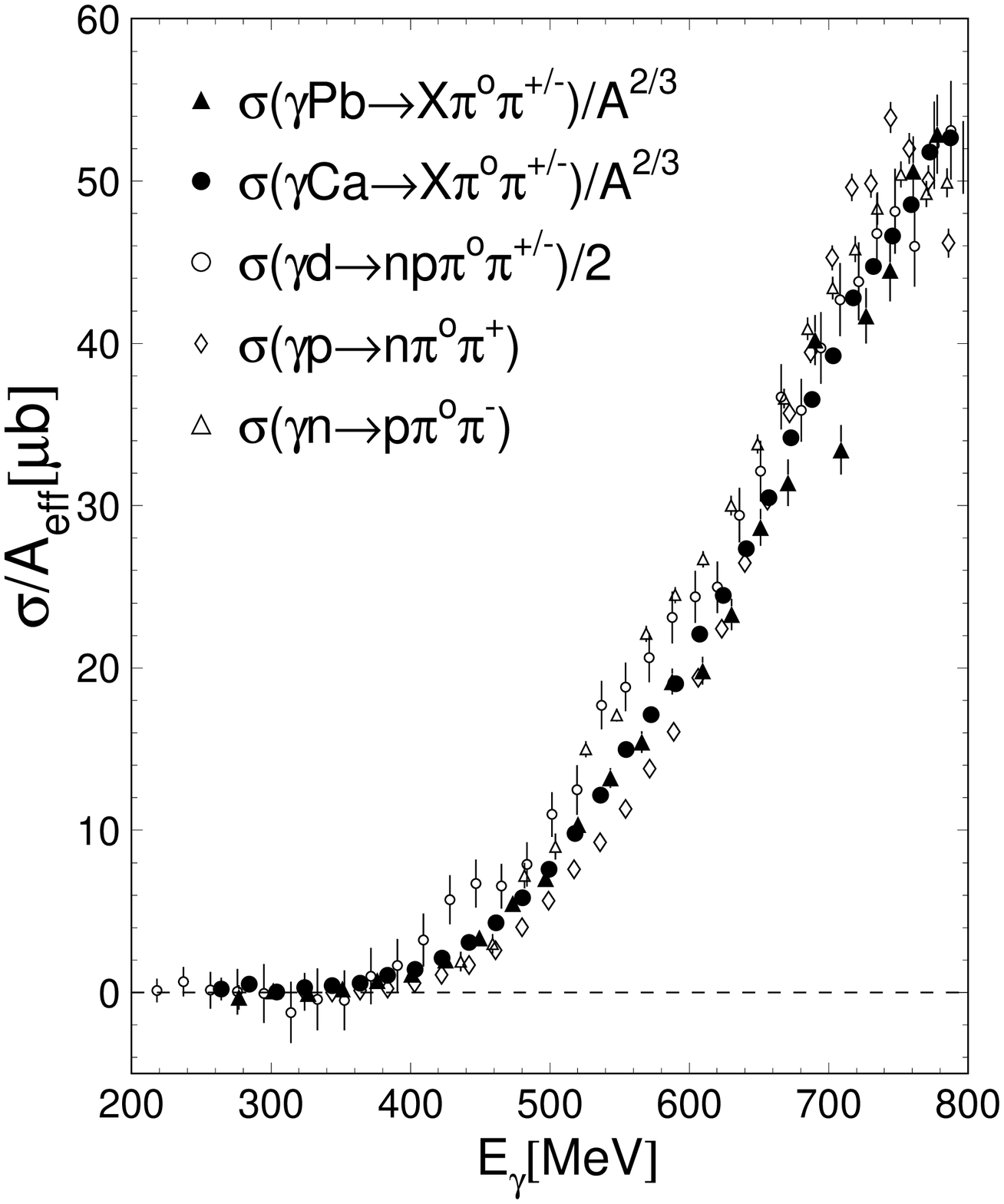}      
}}
\caption{Total cross section of $\pi^0\pi^{\pm}$ 
photoproduction normalized to $A^{2/3}$ and compared to the free nucleon 
cross section (see text). 
}
\label{fig:2pim}       
\end{figure}

The experimental results for double $\pi^0$ photoproduction are summarized in
fig. \ref{fig:2pi0} ($\sigma_{2\pi^0}$). In the first measurement the limited
solid angle did not allow an analysis of events with four photons. Those
data are therefore limited to the practically background free energy range 
below the $\eta$-production threshold (triangles in the figure for C, Ca, Nb, Pb).
The measurement with the improved setup allowed the clean extraction of the 
cross section from 4-photon events via a missing mass analysis. Background
from $\eta$ decays was subtracted. The results for calcium and lead (filled
circles) are shown in the figure up to a maximum photon energy of 800 MeV. 
All nuclear data scale like $A^{2/3}$ and agree with the average nucleon
cross section $\sigma_d/2$ taken from \cite{Kleber_00}. 

The same result is found for $\pi^0\pi^{\pm}$ production which is shown in 
fig. \ref{fig:2pim}. The data for calcium and lead again 
scale like $A^{2/3}$. Also shown in this figure are the cross sections for 
$\gamma p\rightarrow n\pi^0\pi^+$ \cite{Langgaertner_01} and 
$\gamma n\rightarrow p\pi^0\pi^-$ \cite{Zabrodin_97}.
Finally, the cross section for $\gamma d\rightarrow np\pi^0\pi^{\pm}$ was
obtained from the inclusive $\pi^0$ photoproduction cross section
$\sigma_{incl}$ on the deuteron by subtraction of the cross sections from 
all other partial channels (see eq. \ref{eq:incl}). It is obvious from the 
figure that also in this case the nuclear cross sections scaled by $A^{2/3}$
agree with the average nucleon cross section.  
  
The results for the semi-exclusive quasifree reaction channels thus can be 
summarized in the following way: all investigated nuclear cross sections 
are related to a good approximation to the deuteron cross section by:
\begin{equation}
\frac{\sigma_x^{qf}(A)}{A^{2/3}}\approx\frac{\sigma_x^{qf}(d)}{2}
\end{equation}
The scaling among the heavier nuclei holds even when the comparison to the
deuteron is disturbed, e.g. by the effects of coherent single
$\pi^0$ production at low photon energies and or by effects of Fermi smearing
close to the $\eta$ threshold.
The $A^{2/3}$ scaling is the limiting case of strong FSI effects due to the 
short pion mean free path. This means that the quasifree exclusive
reactions probe only the nuclear surface region. Since the properly scaled
nuclear cross sections agree with the deuteron cross section no significant
in-medium effects are observed for the low density surface zone of the nuclei.

\begin{figure}[h]
\resizebox{0.49\textwidth}{!}{%
  \includegraphics{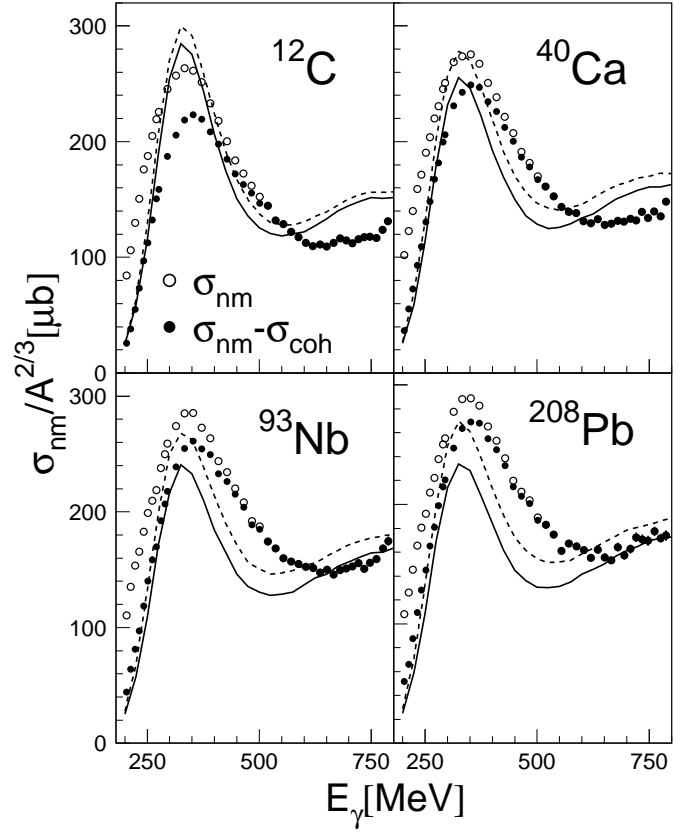}      
}
\caption{Comparison of the total cross section for neutral meson production
($\sigma_{nm}$, see eq. \ref{eq:sigma_cor}) to BUU model calculations. 
Open symbols: $\sigma_{nm}$, filled symbols: coherent single $\pi^0$ production
subtracted. Curves: BUU, $\Delta$ in-medium width from Oset parameterization
(dashed) \cite{Oset_87} or Hirata (solid) \cite{Hirata_79}.} 
\label{fig:sigman}       
\end{figure}

With the knowledge of the nuclear cross sections for $\eta$ and double $\pi^0$
photoproduction it is possible to correct the contribution from higher pion
multiplicity channels to the total inclusive cross section and thus to 
determine the part of the total photoabsorption cross section 
($\sigma_{nm}$, see eq. \ref{eq:sigma_cor}) which corresponds to events 
with at least one $\pi^0$ or one $\eta$ in the final state.
The result is compared in fig. \ref{fig:sigman} to the BUU model calculations.
Again also the results with subtracted coherent part are shown.
The calculations show essentially the same behavior as already observed
for the inclusive cross section in fig. 9. In particular for the lighter nuclei
the second resonance peak is still overestimated and the valley between the 
$\Delta$ resonance and the second resonance region is too pronounced. 
In the case of carbon it should be noted that the assumptions of the BUU model 
are much less fulfilled for such a light nucleus with a relatively large 
surface/volume ratio. However, a further observation is that while for the 
heaviest nuclei the agreement in the second resonance region is much improved, 
the disagreement in the `valley region' around 500 MeV grows. This could mean
that the agreement in the second resonance region is only accidental because 
the contributions from two-body absorption effects are neglected.  
A quantitative investigation of the influence of the two-body absorption 
effects, in particular the dependence on the incident photon energy, is 
therefore highly desirable.

\section{Summary and conclusion}

Cross section data for meson photoproduction from nuclei in the energy range 
from 200 - 800 MeV have been obtained for all final states which involve at 
least one neutral meson. The inclusive data have been used for an analysis of
the pion final state interaction in nuclei. The momentum dependent absorption
properties of nuclear matter show qualitatively the expected behavior:
strong absorption for pions which can excite the $\Delta$ resonance on the
nucleon and almost perfect transparency of the nuclei for low energy pions.
The absorption properties within the BUU model show a similar behavior, 
although no quantitative agreement is achieved for small pion momenta.

The cross sections of all exclusive quasifree meson production reactions
scale with the nuclear surface ($A^{2/3}$) and agree to a good approximation
with the respective cross sections for the deuteron normalized to the deuteron
mass number. This means that no significant in-medium effects are observed
for the quasifree processes which test the low density surface region of
the nuclei. Even the inclusive data, which include some fraction of 
non-quasifree reactions, show a clear signal for the second resonance bump. 
This is an indication that the complete suppression of this bump observed in
total photoabsorption reactions is not the result of trivial nuclear effects
like Fermi motion. The observation that the excitation function for meson
production from the surface region of nuclei has a different energy dependence 
than total photoabsorption in the nuclear volume seems to indicate 
density dependent in-medium effects. 

The experimental data have been compared in detail to calculations in the
framework of the BUU model. It is observed that the model is remarkably 
succesful in reproducing FSI effects such as pion absorption and the shift 
of strength in the pion momentum spectra. However, some discrepancies on 
the quantitative level remain. In the $\Delta$-resonance region the inclusion 
of realistic P$_{33}$ in-medium widths lead to a better description of the 
data. Also in the second resonance region the calculations are close to the 
data although especially for lighter nuclei an overestimation is visible.
The intermediate energy region is not reproduced which
is in line with the theoretical findings in photoabsorption that
multi-body photon absorption mechanisms, which are not included in the BUU 
model, play an important role in this region. Effenberger et al. 
\cite{Effenberger_97b} have calculated such effects for photoabsorption on 
$^{40}$Ca. They find them negligible below incident photon energies of 300 MeV 
and rising to a 25 \% effect at photon energies around 500 MeV. A comparable 
contribution to pion production would much improve the agreement between data 
and calculations in this energy region. However, at present it is not known
how much the two-body effects will contribute at higher incident photon 
energies. A quantitative treatment of these effects is thus necessary 
in future theoretical studies of the in-medium effects in the second
resonance region.

\section{Acknowledgments}
We wish to acknowledge the excellent support of the accelerator group of MAMI,
as well as many other scientists and technicians of the Institut f\"ur
Kernphysik at the University of Mainz. We thank M. R\"obig-Landau for his
contribution to the calibration and analysis of the inclusive data. 
J. Lehr would like to thank M. Post for many valuable discussions.
This work was supported by Schweizerischer Nationalfonds, Deutsche 
Forschungsgemeinschaft, and the UK Engineering and Physical Sciences
Research Council.

\end{document}